\documentclass[12pt]{article}
\usepackage{graphicx}
\begin{document}
\centerline{\bf Ising model simulation in directed lattices and networks}

\bigskip
\centerline{F.W.S. Lima$^1$ and D. Stauffer$^2$}

\bigskip
\noindent
$^1$ Departamento de F\'{\i}sica,
Universidade Federal do Piau\'{\i}, 57072-970 Teresina - PI, Brazil

\medskip
\noindent
$^2$ Instituto de F\'{\i}sica, Universidade Federal Fluminense; 
Av. Litor\^anea s/n, Boa Viagem, 24210-340 Niter\'oi - RJ, Brazil; 
visiting from Institute for Theoretical Physics, Cologne University, 
D-50923 K\"oln, Euroland.

\medskip
e-mail: wel@ufpi.br, stauffer@thp.uni-koeln.de
\bigskip

{\small Abstract: On directed lattices, with half as many neighbours as in the
usual undirected lattices, the Ising model does not seem to show a spontaneous
magnetisation, at least for lower dimensions. Instead, the decay time for 
flipping of the magnetisation follows an Arrhenius law on the square and 
simple cubic lattice. On directed Barab\'asi-Albert networks with two and seven 
neighbours selected by each added site, Metropolis and Glauber algorithms
give similar results, while for Wolff cluster flipping the magnetisation 
decays exponentially with time. 
}

\begin{figure}[hbt]
\begin{center}
\includegraphics[angle=-90,scale=0.5]{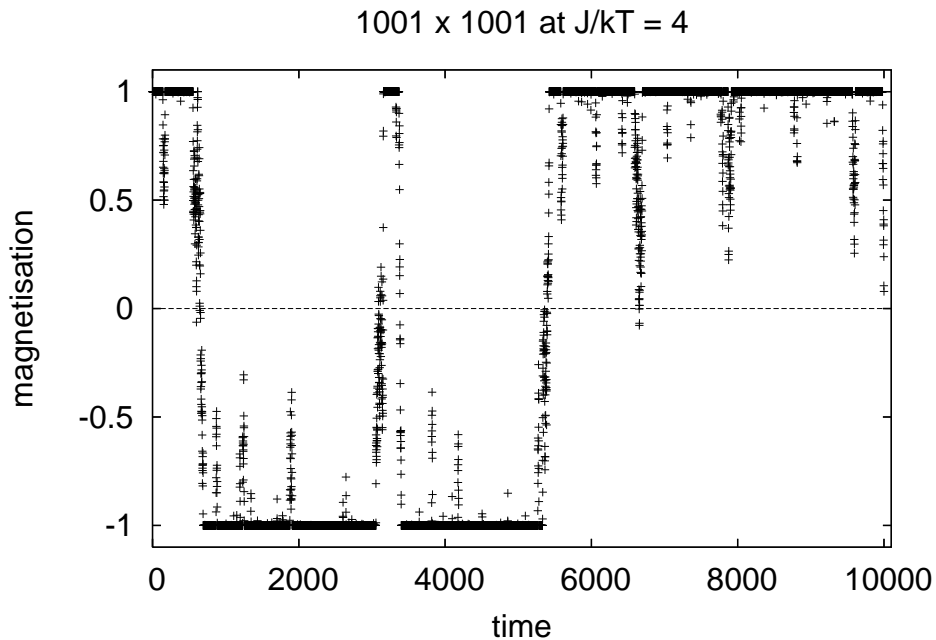}
\end{center}
\caption{
Variation of normalised magnetisation on a square lattice at a temperature 
which is an order of magnitude lower than the critical temperature of the usual 
undirected square lattice.
}
\end{figure}

\begin{figure}[hbt]
\begin{center}
\includegraphics[angle=-90,scale=0.5]{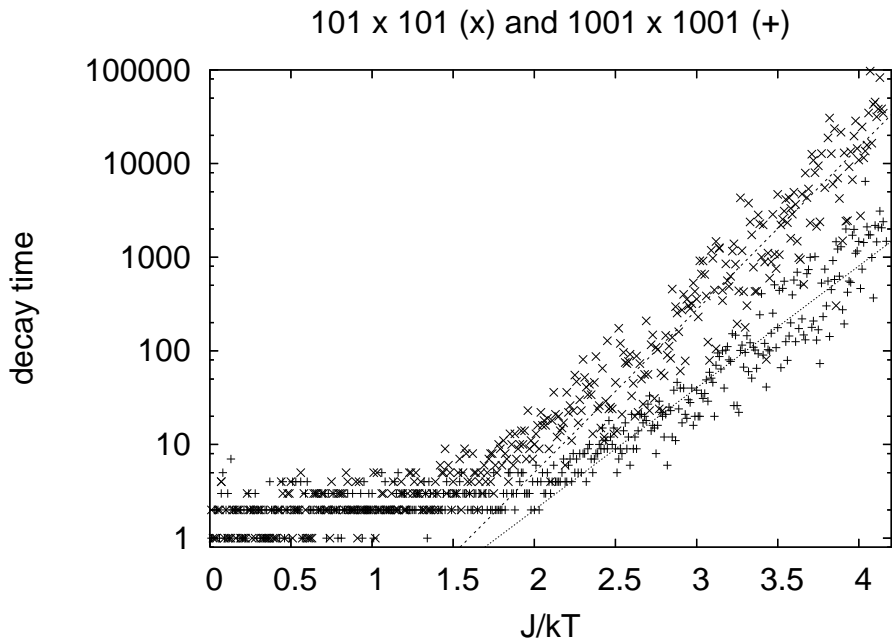}
\end{center}
\caption{
Time for the magnetisation to become negative, versus inverse temperature,
on large (+, lower cloud) and small (x, upper cloud) lattices. We show the 
median of three runs at each temperature. 
}
\end{figure}

\begin{figure}[hbt]
\begin{center}
\includegraphics[angle=-90,scale=0.5]{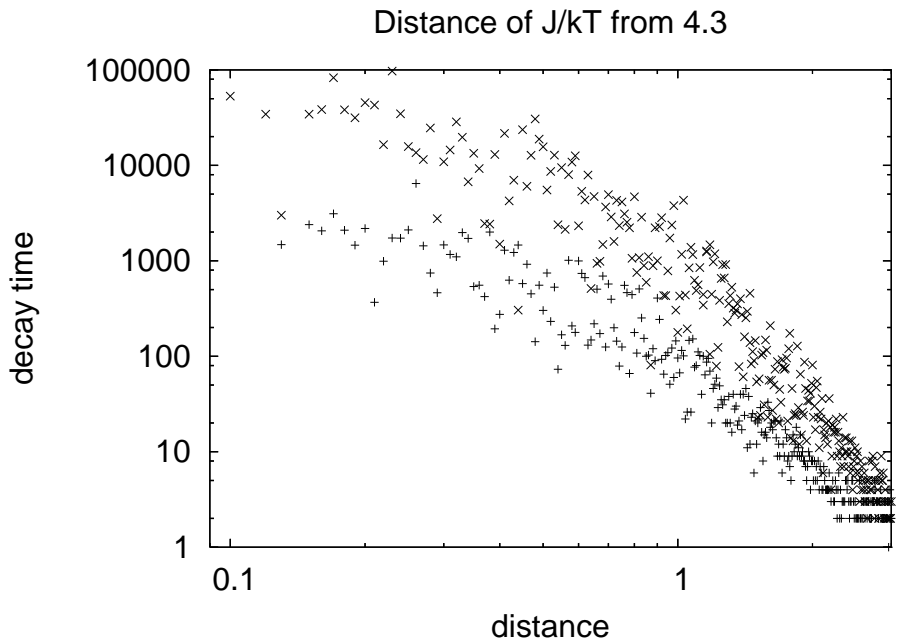}
\end{center}
\caption{
Same data as in Fig.2, shown on a log-log plot with a critical temperature 
assumed at $J/kT = 4.3$. Note that the curvature persists for both small and
large lattices, suggesting that now power law corresponding to a straight
line is valid. 
}
\end{figure}

\begin{figure}[hbt]
\begin{center}
\includegraphics[angle=-90,scale=0.30]{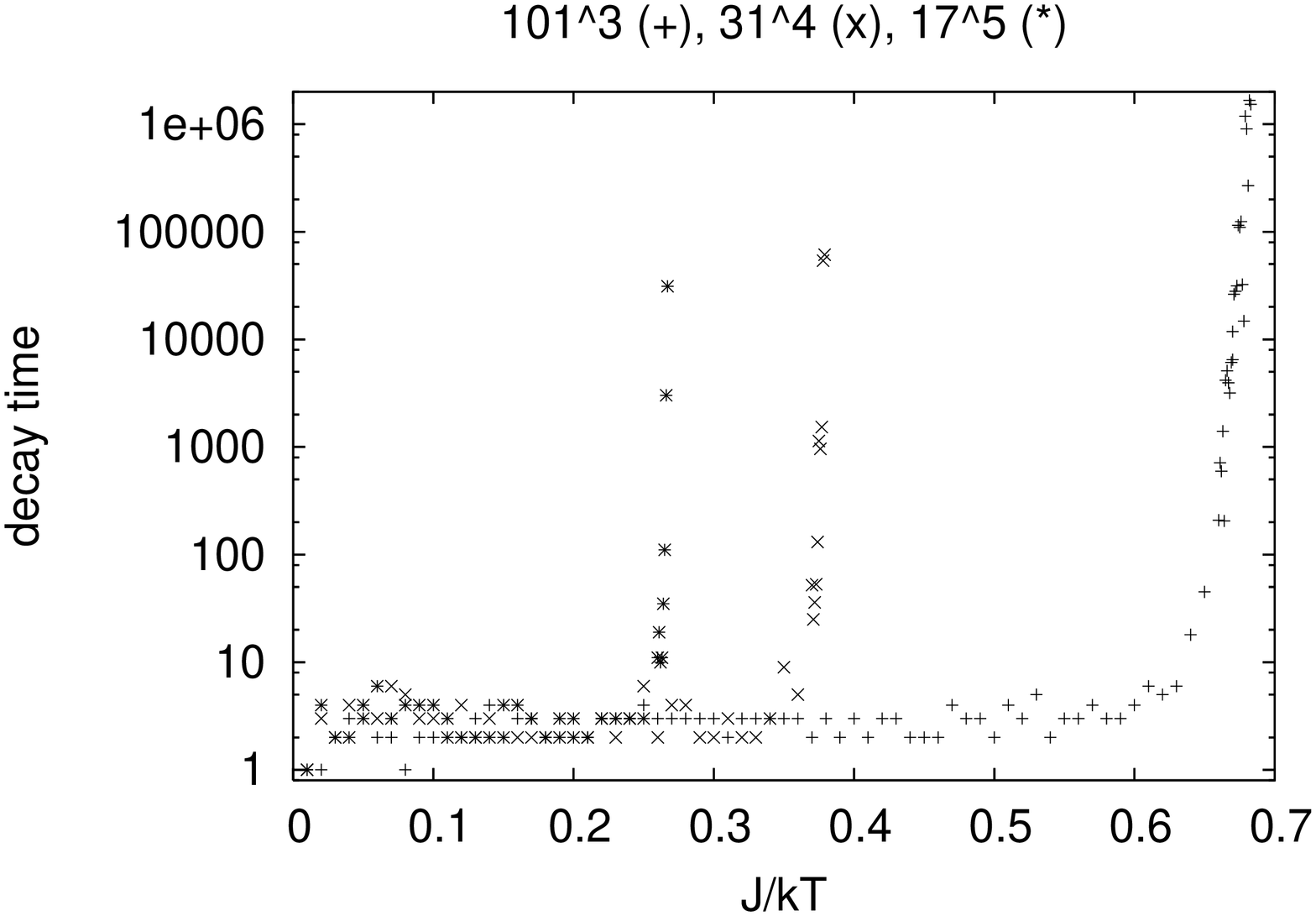}
\includegraphics[angle=-90,scale=0.30]{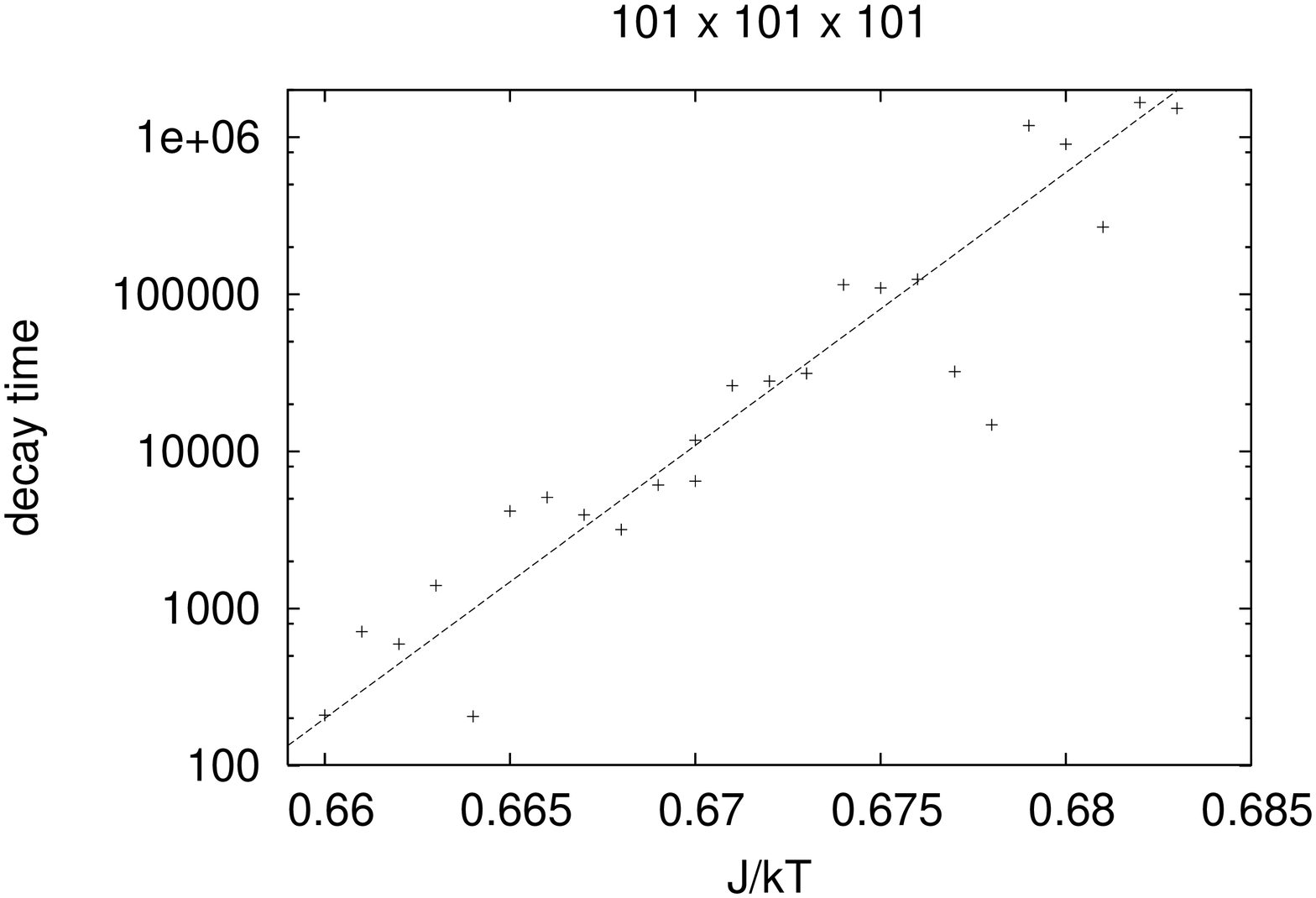}
\end{center}
\caption{
Three, four and five dimensions, plotted analogously to Fig.2 in part a. Part
b shows the three-dimensional data on an expanded horizontal scale, suggesting
also here a straight liner corresponding to an Arrhenius law, time proportional
to exp(Const/$T$). 
}
\end{figure}

\begin{figure}[hbt]
\begin{center}
\includegraphics[angle=-90,scale=0.50]{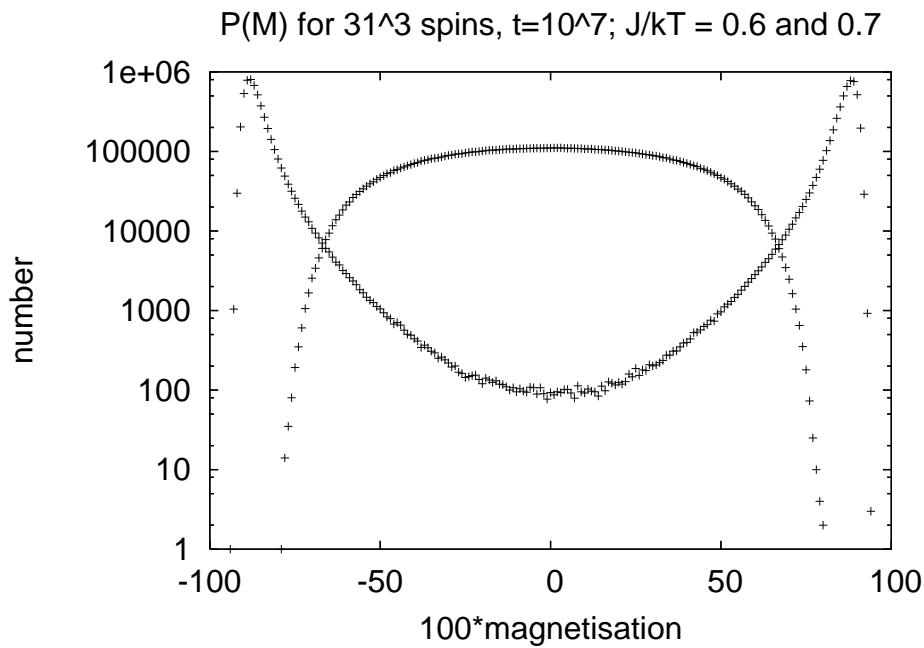}
\end{center}
\caption{
Probability distribution function for the magnetisation (in percent of the
saturation magnetisation) for $31 \times 31 \times 31$  using ten million 
iterations $J/kT = 0.6$ (center) and 0.7 (wings).
}
\end{figure}

\begin{figure}[hbt]
\begin{center}
\includegraphics[angle=-90,scale=0.34]{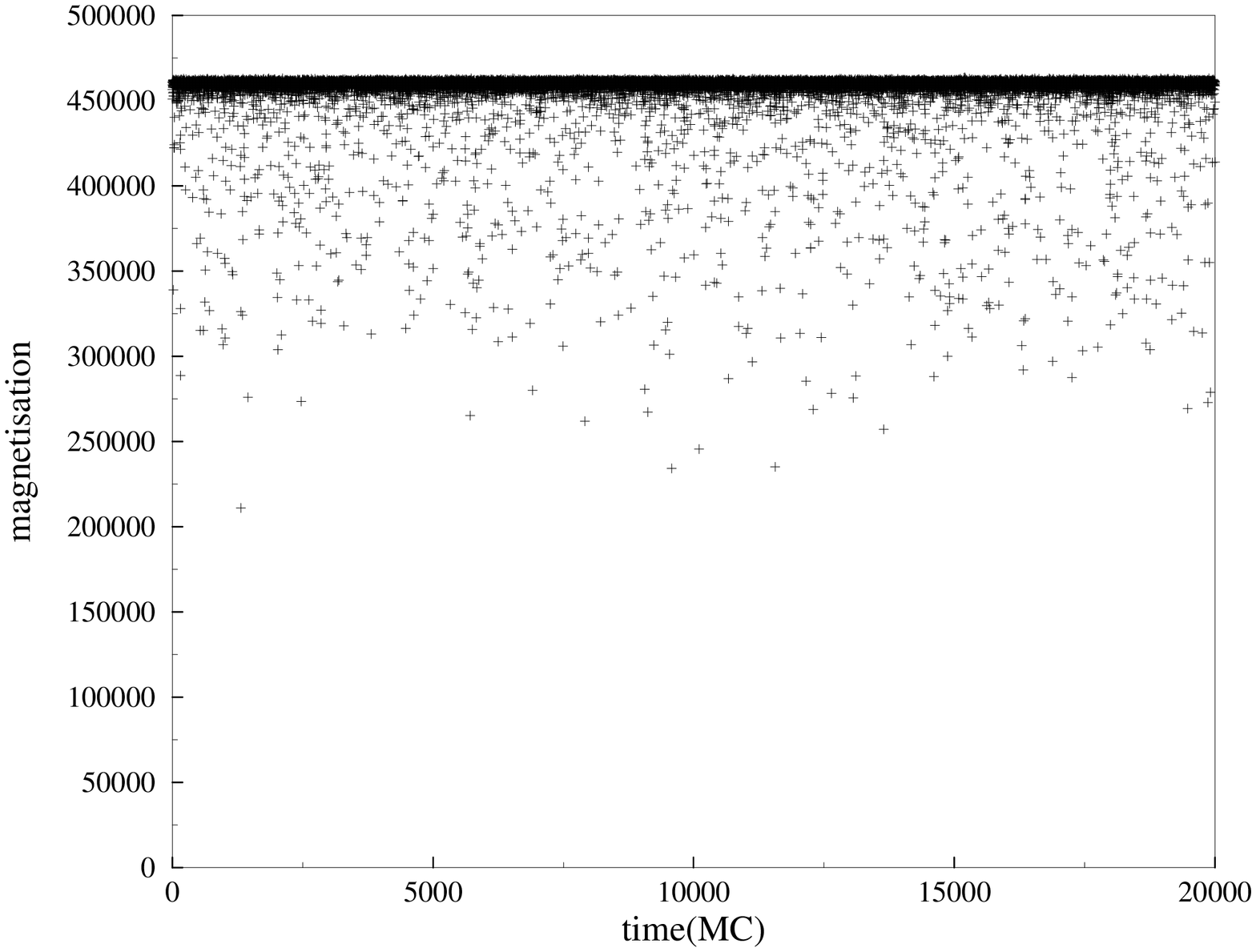}
\includegraphics[angle=-90,scale=0.34]{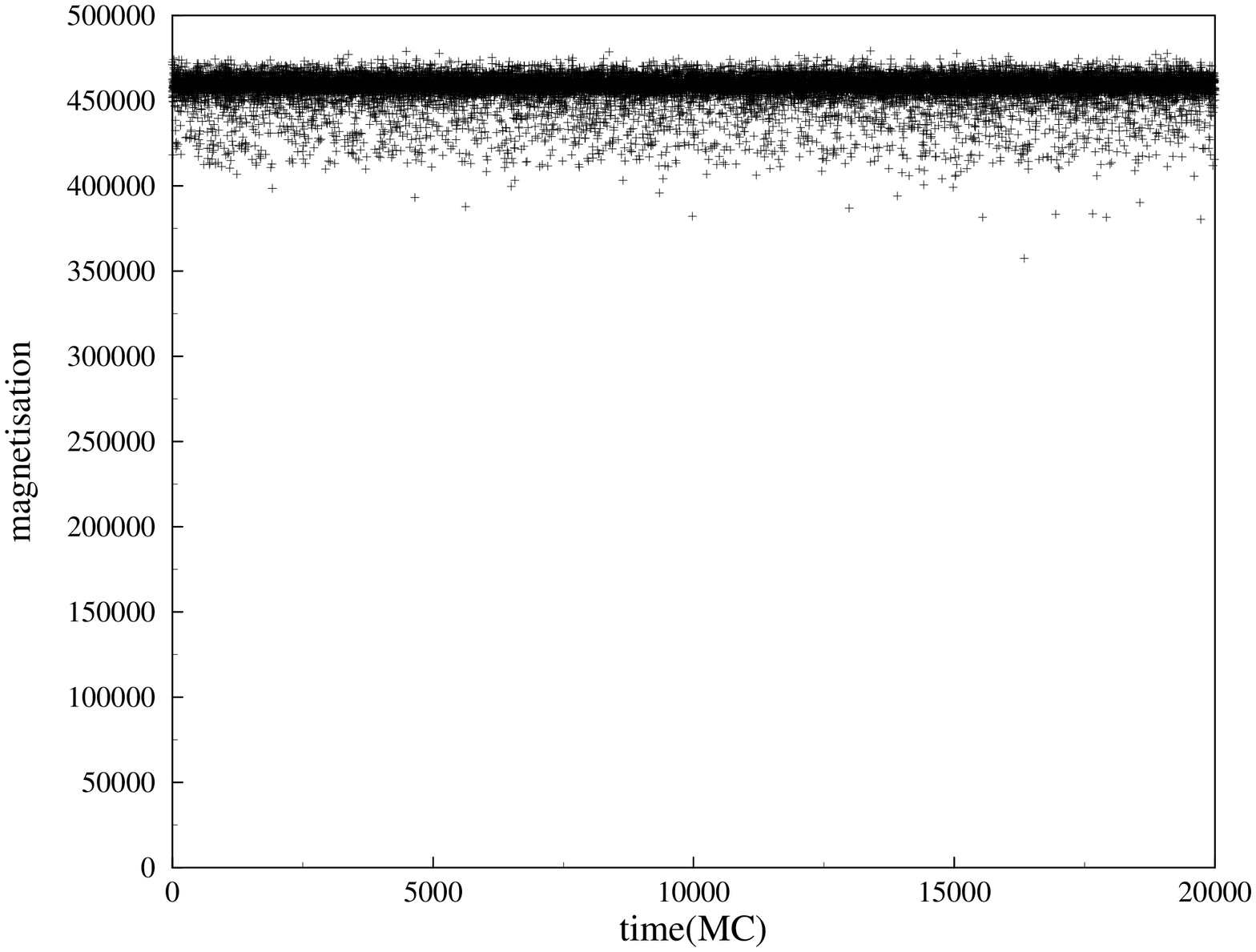}
\end{center}
\caption{
Barab\'asi-Albert network with half a million spins and $m=7$ at $kT/J = 1.7$.
Part a for Glauber kinetics fluctuates more than part b for Metropolis kinetics.
}
\end{figure}

\begin{figure}[hbt]
\begin{center}
\includegraphics[angle=-90,scale=0.50]{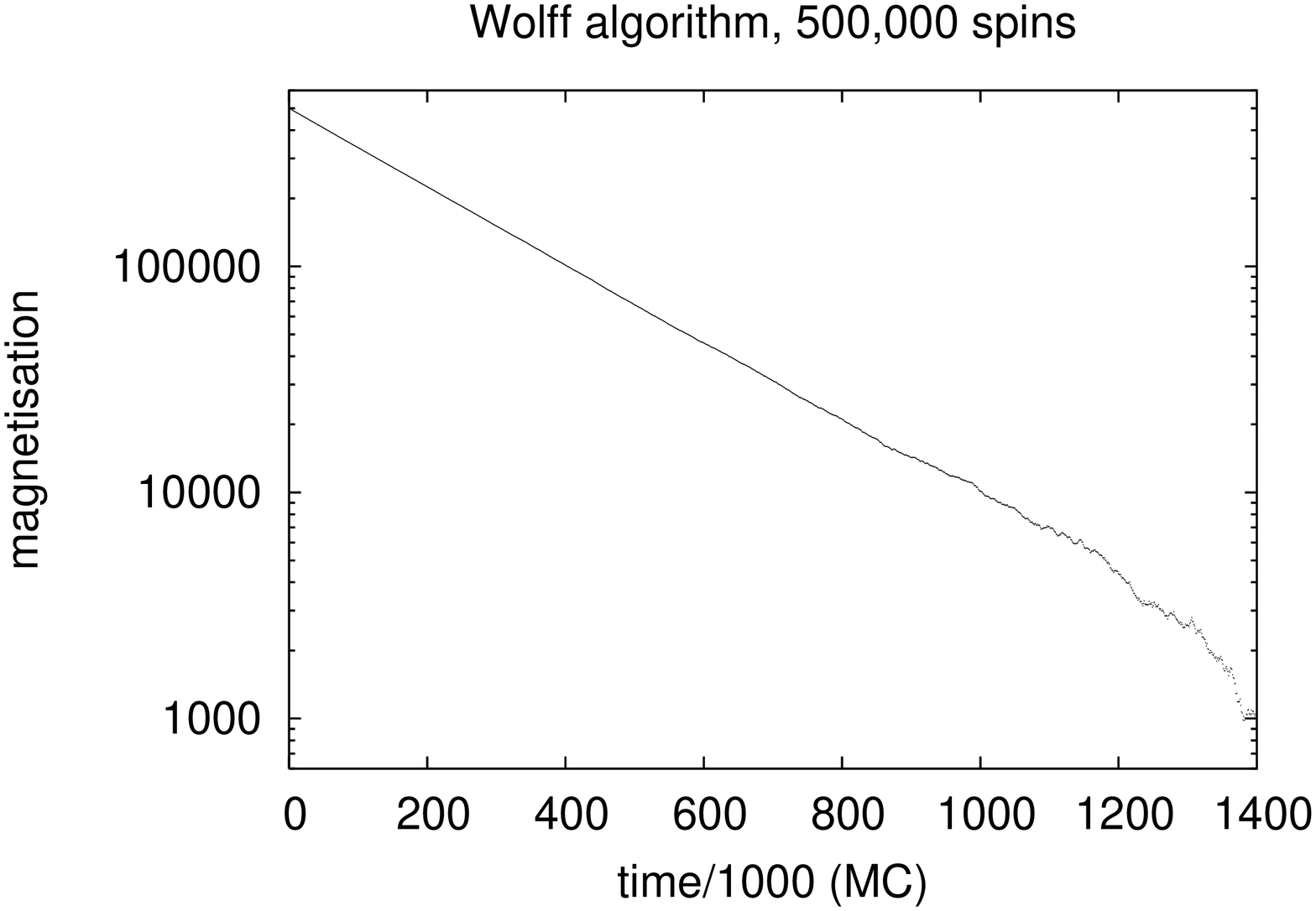}
\end{center}
\caption{
Decay of unnormalised magnetisation in Wolff cluster flip algorithm \cite{wang} 
for same network as in Fig.6, $m=7$ at $kT/J = 1.7$. 
}
\end{figure}

\bigskip

{\bf Introduction}

S\'anchez et al some years ago showed that the Metropolis algorithm, applied to 
directed Watts-Strogatz (small-world) networks has a spontaneous magnetisation, 
even in the limit of directed random graphs\cite{sanchez}. More recently,
Sumour and Shabat \cite{sumour,sumourss} investigated Ising models on 
directed Barab\'asi-Albert networks \cite{ba} with the usual Glauber dynamics. 
No spontaneous magnetisation was 
found, in contrast to the case of undirected  Barab\'asi-Albert networks 
\cite{alex,bianconi} where a spontaneous magnetisation was found below a
critical temperature which increases logarithmically with system size. Now
we simulate directed square, cubic and hypercubic lattices in two to five
dimensions with heat bath dynamics in order to separate the network effects 
form the effects of directedness. And we compare different spin flip algorithms,
including cluster flips \cite{wang},
for Ising-Barab\'asi-Albert networks. In all these cases spins were flipped 
according to algorithms well established to equilibrate systems described by
an Ising energy, even though the directed systems are not described by such an
energy \cite{sanchez}, in contrast to an earlier assertion by one of us
\cite{sumourss}. 

\bigskip
\newpage
{\bf Lattices}

We start with all spins up and update them by going through the lattice 
regularly like a typewriter, numbering them consecutively $i=1,2,3 \dots$.
Helical boundary conditions were used by storing the lowest hyperplane in 
a buffer on top of the lattice; with the upper buffer fixed to +1 the decay 
times were much longer.
The probability for spin $ S_i = \pm 1$ to be +1 in this directed lattice is
$$ p_i = 1/(1 + \exp(2 E_i)/kT), \quad E_i = -J S_i \sum_{k<i} S_k$$
where the inner sum runs over all nearest neighbours $k$ of $i$ with $k < i$,
on the hypercubic lattice; in the usual undirected case the restriction $k < i$
is missing. Thus on a square lattice, the spin $S_i$ is influenced only by its
left and top neighbours, not by the right and bottom neighbours. This directed
influence violates Newton's law $actio = - reactio$. 

For the square lattice, Fig.1 shows an example how the magnetisation jumps
from nearly +1 to nearly --1, similar to \cite{sumourss} and different
from the usual undirected Ising model. Fig.2 shows the time after which the
magnetisation first becomes negative; this time fluctuates strongly but
follows roughly an exponential increase as in an Arrhenius law
$\propto \exp({\rm const} \, J/kT)$. For a lattice with hundred times less spins
the times are slightly larger, suggesting that a rare nucleation event destroys
the metastable state and lets the magnetisation flip. A power law seems to fit
less well, Fig.3.

In higher dimensions, Fig.4, the increase of the decay time happens in a much
smaller interval of $J/kT$ (hundred times smaller in 3D than in 2D)
and perhaps indicates a transition to ferromagnetism.
However, closer inspection of three dimensions at the lowest temperatures,
Fig.4b, suggests also here an exponential Arrhenius increase instead of a
divergence at a positive temperature. Perhaps the same effect happens in four
and five dimensions; and at least in infinite dimensions (or perhaps already
in four dimensions) the slope of log(flipping time) versus $J/kT$ becomes
infinite, meaning a time-independent $T_c$ below which the magnetisation retains
its initial sign. 

Fig.5 shows a probability distribution function $P(m)$ for the magnetisation 
$m$ as in usual Ising models: Two wings for $T < T_c$ and one central peak for 
$T > T_c$. However, in usual large Ising models the magnetization does not 
flip below $T_c$ while in our directed square or cubic lattices it does. The
temperature $T_{\tau}$ above which this probability distribution gets its two 
wings decreases logarithmically towards zero with increasing observation time; 
for $T < T_{\tau}$ the magnetisation retains its initial sign. ($T_{\tau} 
\rightarrow T_c$ for large directed Ising models.) 

\bigskip
{\bf Barab\'asi-Albert Networks}

We now return to the directed scale-free networks of Barab\'asi-Albert type 
simulated by Glauber kinetics in \cite{sumour,sumourss}. We always use half a 
million spins, with each site added to the network selecting $m = 2$ or 7
already existing sites as neighbours influencing it; the newly added spin does
not influence these neighbours. 

With $J/kT = 1.0$ and 1.7, we confirmed \cite{sumour,sumourss} the unusual 
behaviour of the magnetisation, which for $m=7$ stays close to 1 for a long time
inspite of fluctuations. These fluctuations were smaller for Metropolis than
for Glauber (heat bath) kinetics, Fig.6. For $m=2$, the magnetisations are
much smaller (not shown). With Swendsen-Wang cluster flips, the magnetisation
scattered about zero (not shown). Only for Wolff cluster flips, a nice 
exponential decay towards zero is found in Fig.7.

\bigskip
{\bf Conclusion}

In conclusion we found a freezing in of the  magnetisation similar to 
\cite{sumour,sumourss}, following an Arrhenius law at least in low dimensions.
This lack of a spontaneous magnetisation (in the usual sense)
is consistent with the fact
that if on a directed lattice a spin $S_j$ influences spin $S_i$, then spin
$S_i$ in turn does not influence $S_j$. Thus there is no feedback and this 
hinders the stabilisation of a spontaneous magnetisation. The system has
no well defined energy (in contrast to \cite{sumourss}) and no thermal
equilibrium and thus is similar to cellular automata or asymmetric neural 
networks. Thus, even for the same  scale-free networks, different algorithms
give different results. 

We thank P.M.C. de Oliveira for helpful criticism, and J.M. Lopez also for 
drawing our attention on Ref.1 and Fig.3 there.

\end{document}